\documentclass[12pt]{article}
\usepackage{amsmath,amssymb,bm,epsf,epsfig,graphicx,enumerate}
\input epsf.sty
\newcommand{\Tr}{\mathop{\rm Tr}\nolimits}
\newcommand{\re}{\mathop{\rm Re}\nolimits}

\newcommand{\leftpartial}{\mathop{\!\stackrel{\leftarrow}{\partial}}\nolimits}
\newcommand{\rightpartial}{\mathop{\!\stackrel{\rightarrow}{\partial}}\nolimits}

\def\bra#1{\langle #1 |}
\def\ket#1{|#1 \rangle}
\def\aver#1{\left\langle\, #1 \,\right\rangle}

\def\half{\frac{1}{2}}
\newcommand{\VEV}[1]{\left\langle #1\right\rangle}
\newcommand{\msmall}[1]{\mbox{\small$\displaystyle #1$}}
\newcommand{\mscriptsize}[1]{\mbox{\scriptsize$\displaystyle #1$}}

\let\eps = \varepsilon
\def \be {\begin{equation}}
\def \ee {\end{equation}}
\def \bea {\begin{eqnarray}}
\def \eea {\end{eqnarray}}
\def \bdm {\begin{displaymath}}
\def \edm {\end{displaymath}}

\def \rr {{\mathbb R}}

\def \ll {{\cal L}}

\def \aa {{\cal A}}
\def \bb {{\cal B}}

\def \oo {{\cal O}}

\def \vv {{\cal V}}

\def\p {\partial}

\begin{document}
{}~ \hfill\vbox{\hbox{RIKEN-MP-38} }\break
\vskip 2.1cm

\centerline{\large \bf Multibrane Solutions in Open String Field Theory}
\vspace*{8.0ex}

\centerline{\large \rm Masaki Murata$^{a,b,}$\footnote{Email: {\tt mmurata at riken.jp}},
Martin Schnabl$^{a,}$\footnote{Email: {\tt schnabl.martin at gmail.com}}
}

\vspace*{8.0ex}

\centerline {\large \it $^a$ Institute of Physics AS CR, Na Slovance 2, Prague 8, Czech Republic}
\vspace*{2.0ex}

\centerline {\large \it $^b$ Mathematical Physics Lab., RIKEN Nishina Center,}
\vspace*{2.0ex}
\centerline {\large \it Saitama 351-0198, Japan}
\vspace*{2.0ex}

\vspace*{6.0ex}

\centerline{\bf Abstract}
\bigskip
We study properties of a class of solutions of open string field theory which depend on a single holomorphic function $F(z)$. We show that the energy of these solutions is well defined and is given by integer multiples of a single D-brane tension. Potential anomalies are discussed in detail. Some of them can be avoided by imposing suitable regularity conditions on $F(z)$, while the anomaly in the equation of motion seems to require an introduction of the so called phantom term.
\vfill \eject

\baselineskip=16pt


\section{Introduction and summary}

One of the hallmark features of open bosonic string field theory \cite{Witten} is the existence of a tachyon vacuum state, around which there are no perturbative excitations. The perturbative vacuum may describe any consistent D-brane configuration, depending on the choice of boundary conformal field theory (BCFT). As explained by Sen \cite{SenCon}, open bosonic string field theory (OSFT) built upon arbitrary BCFT possesses always another vacuum which corresponds to a state with no D-branes and hence no open string dynamics. Many other solutions have been constructed since then, either numerically or analytically. Some of these solutions describe lower dimensional D-brane configurations. From the viewpoint of the theory expanded around the tachyon vacuum, these solutions correspond to various D-branes popping out of the vacuum, but up to date all solutions found have less energy than the original D-brane. On one hand it might not appear so surprising, as one expects the tachyon condensation to drive the D-brane system to a state with smaller energy. On the other hand, since the final no-D-brane vacuum state is believed to be unique, and string field theory formulated around this vacuum does have solutions with positive energy, it is clear that the apparent impossibility to go higher in the energy is akin to the insufficiency of  a particular coordinate system to describe the whole geometry.

The question is thus, how big is the space of string fields formulated around a given reference BCFT and whether solutions with positive energy with respect to the perturbative vacuum do exist.  This issue has been partially numerically studied in the past by Taylor, Ellwood and also by the second author, using level truncation method, but no conclusive evidence for the existence of such solutions was found. An analytic solution was proposed by Ellwood and the second author, but attempts to compute its energy in level truncation yielded a number that was off by roughly a factor of minus twelve.

In this paper we are going to study a class of universal solutions of the OSFT equations of motion in the form
\be\label{Okawa-sol}
\Psi = F c \frac{K B}{1-F^2} c F,
\ee
where $F=F(K)$, and $K, B, c$ are, by now, well known string fields. These solutions are all universal, in the sense that their form does not depend on the detail of the BCFT. They do exist on any D-brane configuration. Analytic solutions of this form have been studied in \cite{Schnabl,Okawa,FK, Erler1,Erler2} but always with additional assumptions on the function $F$ which more or less guaranteed that the solution was tachyon vacuum. Our proposal is to adopt the least possible assumptions on $F$ and compute the energy\footnote{The energy for a class of tachyon vacuum solutions of this form  with $F(0)=1$ and $F'(0)>0$ was first computed by Erler in \cite{Erler2}. His computations, following quite closely \cite{Schnabl} and being perturbative in $F$, are unfortunately not general enough for present purposes.} and the Ellwood's gauge invariant to deduce the physical meaning of such a solution.

As this paper is rather technical, let us summarize our main assumptions and results. It turns out that the appropriate conditions can be more conveniently stated in terms of a complex function
\be
G(z) = 1- F^2(z).
\ee
The most elementary, but important condition is that $G$ and $1/G$ have inverse Laplace transform which has non-vanishing support only on the positive real axis. This means that we can introduce the function $G(K)$ of a string field $K$
\be
G(K) = \int_0^\infty d\alpha \,  g(\alpha) \, e^{-\alpha K},
\ee
as a superposition of non-negative width wedge states $e^{-\alpha K}$ and hence the star product of such states can in principle be well defined.\footnote{More detailed discussion can be found in \cite{lightning}.} This condition can be restated as the requirement that both $G$ and $1/G$ are holomorphic and that their absolute value is bounded by a polynomial in the half planes $\re z>\eps$ for any $\eps>0$. As we shall see this condition is not very strong and will allow occasionally for some divergences, and in particular it does not guarantee that the energy computed from the action and from the Ellwood invariant will always agree.

Stronger conditions which simplify some computations and give unique answer for the energy can be summarized as follows:
\begin{enumerate}[i)]
\item $G$ and $1/G$ are holomorphic in $\re z \ge0$ except at $z=0$.
\item $G$ is meromorphic at $z=0$.
\item $G$ is holomorphic at $z=\infty$ and has a limit $G(\infty) =1$.
\end{enumerate}
With the help of these assumptions we can evaluate easily the action and we find for the energy $E=\frac{1}{6}\aver{\Psi, Q_B\Psi}$  of the solution (\ref{Okawa-sol})
\be\label{result1}
E=\frac{1}{2\pi^2} \oint_C \frac{dz}{2\pi i} \frac{G'(z)}{G(z)},
\ee
where the closed contour $C$ encircles all singularities and branch cuts in the $\re z<0$ half plane. It does not wind around the origin however. Had some of our assumptions been violated, e.g.  the integrand had branch cuts extending to the boundary of the half-plane, the energy might still be computable, but with appropriate limits taken (if they exist) and additional terms might appear, as we shall discuss in Sec. \ref{sec:Anomalies}.

In our second, independent computation we evaluate the energy using the Ellwood invariant
\be\label{result2}
E = \aver{I|\,c\bar{c} V_{\mathrm{gr}}(i) |\Psi} = -\frac{1}{2\pi^2} \lim_{z \to 0}\, z \frac{G'(z)}{G(z)},
\ee
where $V_{\mathrm{gr}}(i) $ is the properly normalized zero momentum graviton vertex operator inserted at the string midpoint $i$, and $\ket{I}$ is the identity string field.
How can the two expressions (\ref{result1}) and (\ref{result2}) be compatible? Our stronger condition ii) guarantees that the function $G'/G$ is meromorphic in $z=0$ and hence (\ref{result2}) can be written as a tiny contour integral around the origin. At the same time $G'/G$ is holomorphic at infinity and this allows us to deform one contour into the other to prove that both expressions are actually the same!

One notable example of a family of functions obeying the stronger conditions is
\be
G_n(z) = \left(\frac{z}{z+1}\right)^n ,
\ee
for which the energy computed either way is
\be
E = - \frac{n}{2\pi^2} \,.
\ee
For $n=1$ this solution represents the tachyon vacuum solution \cite{ErS}, while $n=0$ gives the perturbative vacuum $\Psi =0$. Negative values of $n$ describe states with energy higher than the perturbative vacuum. In fact, we conjecture that they describe configurations of multiple D-branes. Positive values of $n$ would describe "ghost" branes, objects with negative tension. Such objects are not expected to arise in bosonic string and we show that they are indeed divergent in Fock state expansion.

This paper is organized as follows: In section \ref{Energy-computation} we develop some tools that enable us to compute the energy of our solutions quite efficiently, and we do it in a number of ways. In the following section we discuss possible anomalies when some of our stronger conditions are not met. Section \ref{sec:LevelExp} is devoted to the study of the solutions from level expansion perspective. We will demonstrate that our solutions are well defined in level expansion, nevertheless, we will present arguments for the necessity of adding a so called non-vanishing phantom term. We then test numerically one specific proposal, suggested by several analytic computations, but we do not reach definite conclusions.
Section \ref{Discussion} is devoted to the discussions of various issues touched upon in the main body of the paper.

{\bf Note added:} This is an extended and more detailed version of our conference proceedings report \cite{MS}. As the current paper was nearing the completion, two very interesting papers appeared:
\cite{arXiv:1110.1443,arXiv:1111.2389}. The first one computes the boundary state for the solutions of the type we study, while the second one discusses related results from a broader geometric perspective. Both papers report similar difficulties which we believe are due to not yet fully understood phantom terms.

\section{Computations of the energy}
\label{Energy-computation}
\setcounter{equation}{0}

\subsection{Useful correlators}

To compute the energy using either kinetic or cubic term of the action it is necessary to evaluate number of correlators of the form
\be\label{gen_cor}
\aver{F_1, F_2, F_3, F_4} =
\aver{F_1(K) c F_2(K) c F_3(K) c F_4(K) c B},
\ee
where $F_i(K)$ are ghost number zero string fields, given by arbitrary functions of $K$. In this paper we restrict our attention to the so called geometric string fields in the terminology of \cite{lightning}. This condition means that $F_i$ can be written as formal Laplace transforms
\be
F_i(K) = \int_0^\infty\! f_i(\alpha) \, e^{-\alpha K},
\ee
of arbitrary distributions $f_i$ with support in $[0,\infty)$. One can impose various additional conditions on the distributions as discussed in \cite{lightning}, but in this paper we will formulate them in terms of the properties of the functions $F_i$.

The advantage of such a restriction is twofold: on one hand it gives us a nice geometric picture of the string fields $F_i(K)$ as being superpositions of wedge states $e^{-\alpha K}$ (note that $\alpha$ has to be non-negative), and on the other hand it offers means of computing the correlators through the formula \cite{Schnabl,Okawa,Erler2}
\bea\label{exp_cor}
\VEV{e^{-\alpha_1K}ce^{-\alpha_2K}ce^{-\alpha_3K}ce^{-\alpha_4K}cB}
&=& \VEV{c(\alpha_1)c(\alpha_1+\alpha_2)
c(\alpha_1+\alpha_2+\alpha_3)c(\alpha_1+\alpha_2+\alpha_3+\alpha_4)B}_{C_s}\nonumber\\
&=&  \frac{s^2}{4\pi^3}\Bigg[
\alpha_4\sin\frac{2\pi\alpha_2}s - (\alpha_3+\alpha_4)\sin\frac{2\pi(\alpha_2+\alpha_3)}s\nonumber\\
&&
+ \alpha_2\sin\frac{2\pi\alpha_4}s - (\alpha_2+\alpha_3)\sin\frac{2\pi(\alpha_3+\alpha_4)}s\nonumber\\
&&
+ \alpha_3\sin\frac{2\pi(\alpha_2+\alpha_3+\alpha_4)}s
+ (\alpha_2+\alpha_3+\alpha_4)\sin\frac{2\pi\alpha_3}s\Bigg],\nonumber
\label{eq:cF4}
\eea
which expresses the left hand side as a correlator of four boundary $c$-ghost insertions and one line-integral $b$-ghost insertion on a unit disk presented as a semi-infinite cylinder $C_s$ of width $s$, with the midpoint mapped to infinity. To shorten the notation, we have introduced $s= \sum_{i=1}^4 \alpha_i$.  An explicit formula for the correlator (\ref{gen_cor}) can be obtained from the quadruple integral
\be
\int_0^\infty\!\!\int_0^\infty\!\!\int_0^\infty\!\!\int_0^\infty\! d\alpha_1 d\alpha_2 d\alpha_3 d\alpha_4 \, f_1(\alpha_1) f_2(\alpha_2) f_3(\alpha_3) f_4(\alpha_4) \VEV{e^{-\alpha_1K}ce^{-\alpha_2K}ce^{-\alpha_3K}ce^{-\alpha_4K}c B}
\ee
by a simple trick. Let us insert into the integral an identity in the form
\be\label{sz-trick}
1 = \int_0^\infty ds \, \delta \left(s-\sum_{i=1}^4 \alpha_i \right) = \int_0^\infty ds \int_{-i \infty}^{+i \infty} \frac{dz}{2\pi i}  \,\, e^{sz} \, e^{-z\sum_{i=1}^4 \alpha_i},
\ee
which allows us to treat $s$ as independent of the other integration variables $\alpha_i$. The second equality is just the ordinary Fourier representation of the delta function with the $i$ absorbed in the integration variable, so the contour runs along the imaginary axis. The integrals over  $\alpha_i$ can be easily performed and reexpressed in terms of the original functions~$F_i(z)$
\bea\label{F1234}
\aver{F_1, F_2, F_3, F_4} &=&\int_0^\infty \! ds \int_{-i \infty}^{i \infty} \frac{dz}{2\pi i} \, \frac{s^2}{4\pi^3} e^{s z} \frac{1}{2i} \biggl[- F_1 \Delta F_2 F_3F_4' + F_1 \Delta(F_2 F_3') F_4 \nonumber\\
&& \qquad + F_1 \Delta(F_2 F_3) F_4' - F_1 F_2' F_3 \Delta F_4 +F_1 F_2' \Delta(F_3 F_4) +F_1 F_2 \Delta(F_3' F_4)
\nonumber\\
&& \qquad -  F_1 \Delta(F_2 F_3' F_4) -  F_1 (F_2 \Delta F_3 F_4)' \biggr],
\eea
where for convenience we have omitted common argument $z$ and also introduced an operator $\Delta_s$ defined as
\be
(\Delta_s F)(z) = F\left(z-\frac{2\pi i}{s}\right) - F\left(z+\frac{2\pi i}{s}\right).
\ee
When no confusion arises we omit the subscript $s$. Let us now list some of the properties of this quadrilinear correlator:
\bea
\aver{F_1, 1, F_3, F_4} &=& 0 \nonumber\\
\aver{F_1, F_2, 1, F_4} &=& 0 \nonumber\\
\aver{F_1, F_2, F_3, 1} &=& 0 \nonumber\\
\aver{F_1, K, K, F_4} &=& 0 \nonumber\\
\aver{F_1, F_2, K, K} &=& 0 \nonumber\\
\aver{K, F_2, K, F_4} &=& 0
\label{4corr-prop}
\eea
The first three equations are true because $c^2=0$ and the second two express $cKcKc=0$. These five equations are obeyed at the level of the integrand of (\ref{F1234}) itself.\footnote{The fact that the integrand itself manifestly obeys these five conditions is the reason why we prefer this correlator to the one introduced in \cite{MS}. } The sixth equation can be written as
\be
\aver{Q_B(Bc) F_2 Q_B(c) F_4} = 0,
\ee
and hence it should be true by the basic axiom $\aver{Q_B(\ldots)}=0$. The square-bracket part of the integrand in (\ref{F1234}) for the choice of $F_1=F_3=K$ is equal to
\be\label{eq:KF2KF4}
z (\Delta F_2  F_4 + F_2 \Delta F_4 -  \Delta(F_2 F_4))
- z^2\Delta F_2 F_4' + z \Delta(z F_2) F_4' - z^2 F_2' \Delta F_4 +z F_2' \Delta(z F_4)  -  z (F_2 \Delta z F_4)'.
\ee
The discrete derivative $\Delta_s$ obeys a sort of deformed Leibniz rule
\bea\label{eq:Leibniz}
(\Delta_s f) g + f (\Delta_s g) &=& \Delta_{2s} \left(f\left(z-\frac{\pi i}{s}\right)g\left(z+\frac{\pi i}{s}\right) + f\left(z+\frac{\pi i}{s}\right)g\left(z-\frac{\pi i}{s}\right)\right)
\nonumber\\
&\equiv& \Delta_{2s} (f \circ_s g).
\eea
With the help of this identity (\ref{eq:KF2KF4}) can be rewritten as
\begin{align}
& \Delta_{2s}(F_2 \circ_s zF_4) - \Delta_{2s}(z \circ_s F_2F_4) - \Delta_{2s}(F_2\circ_s z^2F'_4) + \Delta_{2s}(zF_2 \circ_s zF_4') + \phantom{Delta} \nonumber\\
&  +\frac{2\pi i}{s^2} (z\p_z-s\p_s) (-sF_2\Delta_{2s}^2F_4) \,.
\label{eq:KF2KF4Delta}
\end{align}
Now let us consider the $z$-integral in (\ref{F1234}).
If $F_2 \cdot F_4$ is at most $\oo(z^4)$ at infinity, then the integrand is no worse than $\oo(z^{-1})$, and so  we can turn the $z$-integral into the integral over a closed contour $C_s$ by adding a noncontributing arch at infinity in the left half plane Re $z<0$. The subscript $s$ reminds us, that when we try to shrink the closed contour to a finite one, the minimal dimensions of such a contour will typically depend on $s$.
The discrete derivative $\Delta_{2s}$ vanishes under the integral sign
\begin{align}
\oint_{C_s} \frac{dz}{2\pi i} e^{sz} \Delta_{2s}(f_1 \circ_s f_2) = 0 \,,
\label{eq:Delta2s}
\end{align}
for any pair of possibly $s$-dependent functions $f_{1,2}(z)$, provided that the closed contour $C_s$ is sufficiently large so that it encircles all the singularities of $f_i(z)$ and $f_i(z \pm 2\pi i/s)$ in the left half plane Re $z<0$.
We thus arrive at
\begin{align}
\VEV{K,F_2,K,F_4}
= \int_0^\infty ds \oint_{C_s} \frac{dz}{2\pi i} e^{sz} \frac{1}{4\pi^2}
(z\p_z-s\p_s) (-sF_2\Delta_{2s}^2F_4) \,.
\end{align}
The operator $(z\p_z-s\p_s)$ commutes with the exponential $e^{sz}$ and we find that the whole expression reduces to possible surface terms
\begin{align}
\VEV{K,F_2,K,F_4}
&=
\left(\lim_{s\to\infty}-\lim_{s\to0}\right)
\oint_{C_s} dz \, e^{sz} \frac{1}{8\pi^3i}  s^2F_2\Delta_{2s}^2F_4 \,.
\end{align}
The surface term at $s=0$ vanishes if both $F_2$ and $F_4$ are at most $\oo(z)$ at infinity and the one at $s=\infty$ vanishes if $F_2\p^2F_4$ does not have poles on the imaginary axis.

\subsection{Kinetic term}

Let us now describe our computation of the energy using the kinetic term. By a simple manipulation we are led to consider the sum of the following four correlators
\bea\label{psiQpsi}
\aver{\Psi, Q\Psi} &=& \aver{\frac{K}{G}, (1-G), \frac{K}{G}, K G} - \aver{K, (1-G), \frac{K}{G}, K } \\\nonumber && - \aver{\frac{K}{G}, (1-G), K, K } + \aver{K, (1-G), K,\frac{K}{G} }.
\eea
The third term vanishes by \eqref{4corr-prop}. The fourth term can be omitted as well using the stronger assumptions on $G$, but for the sake of generality we shall keep it.
By applying the $s$-$z$ trick (\ref{sz-trick}), the kinetic term can be expressed as
\bea\label{psiQpsisz}
&& \int_0^\infty \! ds \int_{-i \infty}^{i \infty} \frac{dz}{2\pi i} \, \frac{s^2}{8\pi^3 i}\,  e^{s z} \left[\frac{16 \pi i z^2}{s} \frac{G'}{G} - z G \Delta\left(z^2 \frac{G'}{G^2}\right) + 2z \Delta\left( z^2 \frac{G'}{G} \right) + 2z^2 \Delta(z G) \frac{G'}{G^2} \right. \nonumber\\
&& \qquad\qquad\qquad\qquad\qquad \left.
- z \frac{\Delta(z^2 G')}{G} +2 z^2 G' \Delta\left(\frac{z}{G}\right)\right],
\eea
which can be further simplified using the Leibniz rule \eqref{eq:Leibniz}
\begin{align}
\label{eq:EnergyFormula2}
\int_0^\infty ds \int_{-i\infty}^{i\infty} \frac{dz}{2\pi i} \, &\frac{e^{sz}}{8\pi^3 i}
\Bigg[
24\pi i s z^2 \frac{G'}{G}
- 3(z\p_z - s\p_s) \left(s^2 z \frac{\Delta_s (zG)}{G} \right) \\
&
+ 2s^2 \Delta_{2s} \left(z \circ \frac{z^2G'}{G} \right)
- s^2 \Delta_{2s }\left(zG \circ \frac{z^2G'}{G^2} \right)
+ 2s^2 \Delta_{2s} \left(z^2G' \circ \frac{z}{G} \right)
\Bigg].
\nonumber
\end{align}
If $G$ is meromorphic at $z=\infty$, so that it can be written in the form $G=z^r\sum_{n=0}^\infty a_n z^{-n}$, then the square bracket part of the integrand decays as $\oo(1/z^3)$ or faster.
With this condition one can now make the integral along the imaginary axis into a sufficiently large closed contour $C_s$ by adding a non-contributing arch at infinity in the left half plane Re$z<0$,
\begin{align}
\frac{3}{\pi^2} \int_0^\infty ds \oint_{C_s} \frac{dz}{2\pi i} \, & e^{sz}
\Bigg[
s z^2 \frac{G'}{G}
- \frac1{8\pi i}(z\p_z - s\p_s) \left(s^2 z \frac{\Delta(zG)}{G} \right)
\Bigg]\,,
\label{eq:EnergyFormulaCs}
\end{align}
where we have dropped the second line in \eqref{eq:EnergyFormula2} by using \eqref{eq:Delta2s}.

The integration over the remaining two terms can be performed separately.
In the first term the $C_s$ contour can be chosen to be $s$ independent, as the singularities do not depend on $s$. The double integral then makes perfect sense, so the $s$ integration can be performed first\footnote{
If we did not encircle the singularity of $G'/G(z)$ at zero from the left, the double integral would have depended on the integration order.
}
and we find
\begin{align}
\frac{3}{\pi^2} \oint_C \frac{dz}{2\pi i} \frac{G'(z)}{G(z)} \,.
\label{eq:EnergyFirst}
\end{align}
We will see that this value is consistent with the one coming from the Ellwood's gauge invariant observable if $G(z)$ satisfies the conditions i)-iii).

Therefore the second term of (\ref{eq:EnergyFormulaCs}), if nonzero, is an anomaly.
In this term, we can move the $z\partial_z-s\partial_s$ operator outside the exponential factor.
Integrating by parts, only a single surface term can contribute
\begin{align}
- \frac3{\pi^2} \int_0^\infty ds \oint_{C_s}\frac{dz}{2\pi i}
\p_s \left( e^{sz} \frac{izs^3}{8\pi} \frac{\Delta(zG)(z)}{G(z)} \right) \,.
\end{align}
Now, since the contour has been chosen generously large, its infinitesimal variation does not change the integral, and hence one can move the derivative outside the integral, giving rise to the two  surface terms
\begin{align}\label{2surface}
\frac3{\pi^2} \left(\lim_{s\to0} - \lim_{s\to\infty} \right)\oint_{C_s}\frac{dz}{2\pi i}
e^{sz} \frac{izs^3}{8\pi} \frac{\Delta(zG)(z)}{G(z)} \,.
\end{align}
These surface terms could give anomalous contributions so that the value of the kinetic term would be inconsistent with the gauge invariant observable.
Under certain conditions and/or a prescription given in Sec. \ref{sec:Anomalies}, these contributions vanish and we find
\begin{align}
\VEV{\Psi, Q_B\Psi} = \frac{3}{\pi^2} \oint_C \frac{dz}{2\pi i} \frac{G'(z)}{G(z)} \,,
\label{eq:EnergyFormulaC}
\end{align}
which establishes the result (\ref{result1}).
Note that the integral is the well known topological invariant on the space of meromorphic functions, which counts with multiplicity the number of zeros minus the number of poles of $G(z)$ inside the contour $C$. In our case, however, we allow the function to have arbitrary, even essential singularities inside the contour, but nothing outside except for a possible pole or zero at the origin.

One of the crucial assumptions is that both $G(z)$ and $z/G(z)$ are Laplace transforms (corresponding to geometric string fields), and hence $G$ does not have any poles or zeros in the half-plane Re$z>0$. The little bit stronger conditions i)-iii) allow us to shrink the $C$ contour around infinity, picking up only a possible contribution from the origin:
\begin{align}
E = -\frac1{2\pi^2} \oint_{C_0}\frac{dz}{2\pi i} \frac{G'(z)}{G(z)}\,,
\label{eq:energy_zero}
\end{align}
where $C_0$ is now a contour around the origin. There is no contribution from infinity since by assumption $G(z)$ is holomorphic in its neighborhood. For $G(z)\sim z^{-n}$ near the origin we get
\begin{align}
E = \frac{n}{2\pi^2}.
\label{eq:energy_for_n}
\end{align}
Thus the integer $n+1$ can be interpreted as the total number of D-branes on top of each other, including the original D-brane.

At first sight there seem to be no reason why poles of $G$ at the origin would be allowed while zeros not. We will see in Sec. \ref{sec:Convergence} that for $n\leq -4$ the solutions are singular in level expansion. While for the solutions with $n \ge 3$ some of the coefficients are also singular, we believe this is a milder singularity which will be canceled by a phantom term discussed in Sec.~\ref{sec:Phantom}. The solutions with negative value of $n$ have a tension which is too negative and should correspond to the unphysical states with negative number of D-branes. We will call such states as the ghost branes, although this term has been already used in the literature in a rather distinct context.

\subsection{Energy from the cubic term}

The energy of a string field theory solution can also be computed using the cubic term
\begin{align}
E = -\frac16 \VEV{\Psi,\Psi*\Psi}.
\end{align}
Often, this is a useful check, since the solution might not obey the equation of motion automatically when contracted with itself. That one gets the correct answer for the tachyon vacuum in the $\bb_0$ gauge was verified successfully in \cite{Okawa, FK}.

Using the explicit form of the solution (\ref{Okawa-sol}) and reducing the number of $B$ insertions by (anti)commutation we arrive at
\begin{align}
\VEV{\Psi,\Psi*\Psi}
&= \VEV{\frac{K}{G},1-G,\frac{K}{G}(1-G),\frac{K}{G}(1-G)}
- \VEV{\frac{K}{G},1-G,\frac{K}{G},\frac{K}{G}(1-G)^2} \nonumber\\
&
- \VEV{\frac{K}{G}(1-G),1-G,\frac{K}{G}(1-G),\frac{K}{G}}
+ \VEV{\frac{K}{G}(1-G),1-G,\frac{K}{G},\frac{K}{G}(1-G)}.
\end{align}
Now using mere linearity of the four bracket and the identity $KG/G=K$, which is trivially suggested by adopting the $s$-$z$ trick (i.e. it holds at the level of the integrand (\ref{F1234})), we get exactly the same terms as from the kinetic term \eqref{psiQpsi}.
Hence, there is no new computation to be done.


\subsection{Ellwood's gauge invariant observable}
\label{sec:GIO}

We shall evaluate the energy through the gauge invariant observables\footnote{We can evaluate the boundary state following \cite{KOZ}, but for simplicity we will evaluate the gauge invariant observable only.}
\begin{align}
\bra{I} \vv (i) \ket{\Psi}\,,
\end{align}
discovered by Hashimoto and Itzhaki \cite{HI} and independently by Gaiotto, Rastelli, Sen and Zwiebach \cite{GRSZ}. They depend on an on-shell closed string vertex operator $\vv = c\bar c\, V^{m}$ inserted at the string midpoint $i$ in the upper-half-plane coordinates. The state $\bra{I}$ is the identity string field and it also plays the role of the Witten's integration.
Ellwood conjectured \cite{Ellwood} that the gauge invariant observables compute the difference of the one-point functions of the closed string on the unit disk between the trivial vacuum and the one described by the solution $\Psi$:
\begin{align}
\bra{I} \vv(i) \ket{\Psi}
= \aa^{\rm disk}_\Psi(V^{\rm m}) - \aa^{\rm disk}_0(V^{\rm m})\,.
\end{align}
Here $\aa^{\rm disk}_\Psi(V^{\rm m})$ is the one-point function of the matter part of the closed string vertex operator $V^{\rm m}$ on the disk in the vacuum $\Psi$.
For the tachyon condensation solution, Ellwood confirmed that the one-point function is zero.
For the $N$-brane solution the one-point function is expected to be $N$ times the one in the trivial vacuum:$\aa^{\rm disk}_\Psi(V^{\rm m})= N \aa^{\rm disk}_0(V^{\rm m})$.

Now let us compute the gauge invariant observable for the solution \eqref{Okawa-sol}.
Since $K$ commutes with the operator inserted at the midpoint, the invariant equals
\begin{align}
\VEV{\vv\, FcB\widetilde{F}cF}
= \VEV{\vv\, cB\widetilde{F}cF^2},
\end{align}
where $\widetilde{F}=K/(1-F^2)$ and $\vv=\vv(i)\ket{I}$.
When we express $F^2$ and $\widetilde{F}$ as the Laplace transform and denote the inverse Laplace transforms as $f$ and $\widetilde{f}$ respectively, it becomes
\begin{align}\label{Elw-corr}
\int_0^\infty d\alpha \int_0^\infty d\beta \, \widetilde{f}(\alpha) f(\beta)
\VEV{\vv\, cBe^{-\alpha K} c e^{-\beta K} }\,.
\end{align}
A very similar correlator has been computed by Ellwood \cite{Ellwood}, so with the help of a simple reparametrization we find
\begin{align}
\int_0^\infty d\alpha \int_0^\infty d\beta\, \widetilde{f}(\alpha) f(\beta)
\frac{2i}{\pi} \beta \VEV{V^{\rm m}}^{\rm matter}_{\rm UHP}
&= \lim_{z, w\, \to0}\widetilde{F}(z)\, \partial_w F^2(w) \, \aa^{\rm disk}_0(V^{\rm m})\nonumber\\
&= - \left(\lim_{z\to0}z\frac{G'(z)}{G(z)}\right) \aa^{\rm disk}_0(V^{\rm m}) \,,
\label{Elwab}
\end{align}
where $\aver{\cdot}^{\rm matter}_{\rm UHP}$ is the matter correlator on the upper half plane (UHP) and $G=1-F^2$.
In general $\partial_z F^2(z)$ or $\widetilde{F}(z)$ may diverge in the $z \to 0$ limit, but their product is much better behaved. There is a natural regularization which makes this manifest. Imagine replacing string field $K$ with $K+\eps$ in the solution \eqref{Okawa-sol}, where $\eps$ is a small but positive number (multiplied by the identity string field). The only change to \eqref{Elw-corr} is the appearance of an extra factor of $e^{-(\alpha+\beta)\eps}$ under the integral sign. After evaluating the correlator we get $\widetilde{F}(\eps) \partial F^2(\eps)$ and in the limit we justify the final expression in \eqref{Elwab}.

Substituting the result \eqref{Elwab} into the Ellwood's relation we get
\begin{align}\label{Elwres}
\aa^{\rm disk}_\Psi(V^{\rm m}) = \left(1- \lim_{z\to0}z\frac{G'(z)}{G(z)}\right) \aa^{\rm disk}_0(V^{\rm m}) \,.
\end{align}
Now if $G(z)$ behaves as $z^{-n}$ around the origin,
\begin{align}
\aa^{\rm disk}_\Psi(V^{\rm m}) = (n+1) \aa^{\rm disk}_0(V^{\rm m}).
\end{align}
Thus the solution $\Psi$ can be expected to describe $n+1$ copies of the original D-brane. In particular, one can compute the energy from the Ellwood invariant by using appropriately normalized zero momentum graviton vertex operator
\be
E = \aver{I|\,c\bar{c} V_{\mathrm{gr}}(i) |\Psi} = -\frac{1}{2\pi^2} \lim_{z \to 0}\, z \frac{G'(z)}{G(z)}.
\ee
For a function $G$ meromorphic at the origin, or more generally one for which $G'/G$ is meromorphic, the limit can be replaced by a contour integral
\be
\lim_{z \to 0}\, z \frac{G'(z)}{G(z)} = \oint_{C_0} \frac{dz}{2\pi i} \, \frac{G'(z)}{G(z)},
\ee
where $C_0$ is a small contour encircling the origin. As we have explained in the introduction, this result is consistent with our energy computation using the kinetic term, if the conditions i)-iii) hold.

Let us now show how the result \eqref{Elwres} can be obtained directly in the contour form using the $s$-$z$ trick. This exercise will hopefully help the reader to acquire more familiarity with the trick itself. By the rules of the trick, the left hand side of \eqref{Elwab} can be written as
\begin{align}\label{NormElw}
\frac{\bra{I} \vv(i) \ket{\Psi}}{\aa^{\rm disk}_0(V^{\rm m})}
= - \int_0^\infty ds \int_{-i\infty}^{i\infty} \frac{dz}{2\pi i}\, e^{sz} \, z\frac{G'}{G}\,.
\end{align}
Suppose that $G=z^r\widetilde{g}(1/z)$ with $\widetilde{g}$ holomorphic at the origin.
Since $zG'/G$ behaves as $\oo(z^0)$ around infinity (for $r \ne 0$), we cannot simply close the line $z$-integral into a closed contour, but we have to add a delta function $\delta(s)$. In total we have
\begin{align}
\frac{\bra{I} V_c(i) \ket{\Psi}}{\aa^{\rm disk}_0(V_c^{\rm m})}
= - \int_0^\infty ds \oint_{C} \frac{dz}{2\pi i} \, e^{sz} \, z\frac{G'}{G} - r
= \oint_{C} \frac{dz}{2\pi i} \frac{G'}{G} - \oint_{C_\infty} \frac{dz}{2\pi i} \frac{G'}{G}
= - \oint_{C_0} \frac{dz}{2\pi i} \frac{G'}{G},
\end{align}
where $C_\infty$ is the circle with sufficiently large radius encircling all the poles of $G'/G$.
In the last equation, we have assumed the condition i).

To close the discussion let us see what happens when we combine the $s$-$z$ trick with the $K+\eps$ regularization. The normalized Ellwood invariant \eqref{NormElw} becomes
\be\label{Kesz}
\frac{\bra{I} \vv(i) \ket{\Psi}}{\aa^{\rm disk}_0(V^{\rm m})}
= - \lim_{\eps\to 0} \int_0^\infty ds \int_{-i\infty}^{i\infty} \frac{dz}{2\pi i}\, e^{sz} \, (z+\eps)\frac{G'(z+\eps)}{G(z+\eps)}\,.
\ee
Assuming for simplicity our condition iii), i.e. $G(\infty)=1$, the integrand decays rapidly in the $\re z<0$ half-plane and therefore the contour can be closed. The pole (or zero) of $G$ at the origin shifted to $-\eps$ does not contribute to the integral, because of the factor $z+\eps$. The contour can be moved left,  so that the point $-\eps$ is just to the right of the contour.
The resulting double integral is now absolutely convergent, so the $s$-integral can be performed first. We find
\be
\frac{\bra{I} \vv(i) \ket{\Psi}}{\aa^{\rm disk}_0(V^{\rm m})}
=  \lim_{\eps\to 0}  \oint_{C} \frac{dz}{2\pi i} \frac{z+\eps}{z} \frac{G'(z+\eps)}{G(z+\eps)} =\lim_{\eps\to 0}  \oint_{C} \frac{dz}{2\pi i} \frac{z}{z-\eps} \frac{G'(z)}{G(z)} = \oint_{C} \frac{dz}{2\pi i} \frac{G'}{G}.
\ee
In the second equality we have shifted the variable $z$ to $z-\eps$ and correspondingly the contour by $\eps$ to the right. In the last equality we took the limit, noting that neither $z=0$ nor $z=\eps$ contribute, as they lie outside the contour.

Notice that the invariant comes entirely from the $\eps$ term in the factor $(z+\eps)$ in \eqref{Kesz}. Had we considered only the first $z$ term\footnote{This would be actually equivalent to computing the Ellwood invariant for a pure gauge solution with $G(K+\eps)$.}, we would not have been allowed to move the contour past $-\eps$, and the total contribution would have been zero. This is quite a general feature of Ellwood invariants that they receive contribution only from a naively vanishing piece, a so called phantom term. In Sec.~\ref{sec:Phantom} we will propose another type of a non-vanishing phantom, associated to the residue at $z=0$ in the $s$-$z$ trick computation. This term does not contribute to the Ellwood invariant at finite $\eps$. But it does when we set $\eps=0$ strictly at the beginning, in the sense that the contribution from the residue at the origin must be subtracted (or not counted) in the solution.


\section{Possible Anomalies}
\label{sec:Anomalies}
\setcounter{equation}{0}

In the formalism of string field theory one occasionally encounters expressions which are either divergent or anomalous and must be treated with due care. One of the least understood type of divergences is associated to the so called sliver divergence. The string fields $e^{-\alpha K}$ describe wedge states of width $\alpha$, but these do not vanish in the large $\alpha$ limit, contrary to what one would naively expect from the large $\alpha$ behavior of $e^{-\alpha K}$ if $K$ was thought to be positive definite. Such anomalies have been recently studied by Erler and Maccaferri \cite{EM} (cf. \cite{BGT1,BGT2} for an alternative viewpoint), following an inspiring work on tachyon lumps by Bonora et al. \cite{BMT}. In our computation of the energy or the Ellwood's invariants, we have repeatedly used the Laplace representation of various functions of $K$, without really worrying whether they exist as string fields. This has allowed us to make fast computational progress but left behind many questions. In our computation of the energy using the kinetic term or the Ellwood's invariant we have found two potentially conflicting results. We shall study the surface terms \eqref{2surface} and look under what conditions they vanish. In the last subsection we will examine possible anomaly in the equations of motion.

\subsection{Anomaly at $s=\infty$}

Let us look at the $s=\infty$ surface term \eqref{2surface} in the kinetic term \eqref{psiQpsi}
\be\label{sinfty}
-\lim_{s \to \infty} \frac{3}{\pi^2} \oint_{C_s} \frac{dz}{16 \pi^2} \, e^{ z s} \, z s^3 \frac{\Delta (z G)}{G(z)}\, .
\ee
If the singularities or zeros of $G$ are all at $\re z<0$, then the contour $C_s$ can be deformed, moved away to the left of the imaginary axis, and the suppression factor $e^{sz}$ would guarantee that the $s\to\infty$ limit is vanishing. For fractional branes\footnote{These singular solutions should not be confused with fractional D-branes on orbifolds.}, however, there is a branch cut in $G$ going all the way to zero. One can consider for example
\be
G(z) = \left(\frac{z+1}{z}\right)^r,
\ee
where for non-integer $r$, the branch cut might be chosen to run along the real axis from $-1$ to $0$.
In such a case one can suspect that the $s =\infty$ surface term is non-vanishing, which indeed happens, as we will now show. Let us write
\be
G(z) = \frac{1}{z^r} g(z),
\ee
assuming that $g(z)$ is holomorphic and nonvanishing at zero. The contour integral in \eqref{sinfty} is well defined for $ -2 < r < 2$. For non-integer $r$ outside this range, the contour integral is still finite,  with the prescription of encircling the branch point at $z=0$ from the right. After a change of variables $ z \to z/s$ the integrand possesses well defined holomorphic limit for large $s$. Closing the integration contour and further changing the integration variable to $ w=1/z$ we find the anomaly
\bea\label{Anomalysinfty}
A_{s=\infty} &=&  -\frac{3}{\pi^2} \oint \frac{dw}{16\pi^2} e^{1/w} \frac{1}{w^4} \left(  (1-2\pi i w)^{1-r} - (1+2\pi i w)^{1-r}\right) \nonumber\\
&=& -\frac{1}{2} (r^3-r) \left[ {}_1 F_1(2+r,4,2\pi i)+{}_1 F_1(2+r,4,-2\pi i) \right].
\eea
This anomaly vanishes only at $r=-1, 0, 1$ and at a sequence of non-integer values $r= \pm 3.63948, \pm 8.12955, \pm 14.2036, \ldots$. At these latter values the surface term vanishes, but generically we expect the solutions to be still singular. But what about integer values $|r| \ge 2$? If the anomaly was unavoidable, we would have to conclude that the multiple brane solutions (beyond the double-brane solutions) do not exist. For integer values of $r$, one has the option to take the $z$ contour to bypass the singularity at the origin from the left.\footnote{Note that the integrand of the first term in (\ref{eq:EnergyFormulaCs}) is regular at zero, so the problem has not arisen in our previous discussion.}
We will see that the same prescription is needed also to avoid anomaly in the equations of motion. This prescription will be interpreted in Sec. \ref{sec:Phantom} as an extra phantom term visible in level expansion. For non-integer brane with $|r|>2$ this option is not possible since one cannot bypass non-integrable singularity on the side of a branch cut.
Therefore we expect that the equation of motion should not hold if $r$ is fractional and so the fractional brane solutions do not exist.

If $G$ had additional zeros or poles on the imaginary axis, except at the origin, and if we encircled them from the right, these would make the surface term \eqref{sinfty} behave as $\oo(s^2)$, i.e. quadratically divergent.\footnote{Sometimes an infinite number of zeros or poles on the imaginary axis might conspire to give vanishing contribution. This happens for example for the tachyon vacuum described by the function
$G(z)=1-e^{-z}$.}
Therefore, one would conclude that the contour $C_s$ should encircle them from the left.
However, if we do so, the value of the kinetic term becomes incompatible with the Ellwood invariant since the zeros and poles on the imaginary axis produce anomaly if we deform the contour $C$ into $C_0$.


\subsection{Anomaly at $s=0$}

Identical computation can be done to study under what conditions does the $s=0$ surface term contribute. Assuming that the behavior of $G$ around $z=\infty$ is of the form
\be
G(z) = \frac{1}{z^r} \tilde g(1/z),
\ee
with $\tilde g$ holomorphic around zero, one finds exactly the same anomaly (\ref{Anomalysinfty}) up to a minus sign due to the overall sign in front of the $s=0$ contribution:
\be\label{Anomalys0}
A_{s=0}
= \frac{1}{2} (r^3-r) \left[ {}_1 F_1(2+r,4,2\pi i)+{}_1 F_1(2+r,4,-2\pi i) \right].
\ee
The anomalies $A_{s=0}$ and $A_{s=\infty}$ may cancel if the power-law behavior at infinity and at zero is the same. To get rid of the anomalies in the equations of motion (discussed in the next section), one has to pick a contour bypassing zero on the left and then $A_{s=\infty}$ will automatically vanish. The anomaly $A_{s=0}$ will vanish independently due to the condition iii), i.e. $\lim_{z\to\infty} G(z) =1$, which we impose anyway.

\subsection{Anomalies in the equations of motion}
\label{s_anom_eom}

Let us take the simplest candidate multiple brane solution and regularize it by replacing  $K$ with $K+\eps$
\be
\Psi =  c \, \frac{K+\eps}{G(K+\eps)} \, B c \, (1-G(K+\eps)).
\ee
This regularized solution can be viewed as a sum of a pure gauge string field with $G(K+\eps)$, plus an additional piece proportional to $\eps$. In Sec. \ref{sec:GIO} we have seen that this naively vanishing term provides the full contribution to the Ellwood invariant. This term is reminiscent of the $\psi_N$ piece of \cite{Schnabl} which have been dubbed a phantom term, since it was vanishing in level expansion, but nevertheless it contributed non-trivially to the energy.
Here, the term linear in $\eps$ is equally important, and vanishes in level expansion as well. We shall call it the $\eps$-phantom, to distinguish it from the finite phantom studied in Sec.~\ref{sec:Phantom}.

Thanks to the $\eps$-phantom, the equation of motion now reveals a possible anomaly
\bea
\aa_\eps &=& Q\Psi + \Psi*\Psi \nonumber\\
&=& \eps c \frac{K+\eps}{G(K+\eps)} c (1-G(K+\eps)),
\eea
and the important question is whether it vanishes as $\eps$ is sent to zero or not.
We shall explicitly check its first coefficient in the level expansion.

To obtain the coefficient $t_{F1F2F3}$ in front of $c_1 c_0 \ket{0}$ in a general expression $F_1 c F_2 c F_3$ we need first the elementary formula
\bea
e^{-\alpha K} \, c \, e^{-\beta K} \, c \, e^{-\gamma K} &=& \hat U_{1+\alpha+\beta+\gamma} \tilde c \left(\frac{\pi}{4}(-\alpha+\beta+\gamma)\right)\tilde c \left(\frac{\pi}{4}(-\alpha-\beta+\gamma)\right) \ket{0} \nonumber\\
 &=& t_{\alpha\beta\gamma} c_1 c_0 \ket{0} + \cdots,
\eea
where
\be
t_{\alpha\beta\gamma} = - \left(\frac{s+1}{2}\right)^3 \left(\frac{2}{\pi}\right)^2 \sin\left(\frac{\pi}{2} \frac{1+2\alpha}{s+1}\right)
\sin\left(\frac{\pi}{2} \frac{2\beta}{s+1}\right)
\sin\left(\frac{\pi}{2} \frac{1+2\gamma}{s+1}\right).
\ee
With the help of the standard $s$-$z$ trick we readily find
\bea\label{tF123}
t_{F1F2F3} &=& \left(\frac{2}{\pi}\right)^2 \frac{1}{8i} \int_0^\infty\! ds \int_{-i \infty}^{i \infty} \frac{dz}{2\pi i} \, e^{ z s} \left(\frac{s+1}{2}\right)^3 \left[F_1(z-\omega) e^{\frac{\omega}{2}} - F_1(z+\omega) e^{-\frac{\omega}{2}} \right]
\nonumber\\
&& \qquad \times \bigl[F_2(z-\omega)  - F_2(z+\omega) \bigr]
\left[F_3(z-\omega) e^{\frac{\omega}{2}} - F_3(z+\omega) e^{-\frac{\omega}{2}} \right],
\eea
where
\be
\omega = \frac{\pi i}{s+1}\,.
\ee
Setting $F_1 = 1$, $F_2 =\frac{z+\eps}{G(z+\eps)}$ and $F_3 = 1-G(z+\eps)$, where $G$ is holomorphic in $\re z<0$ with no zeros or singularities on the imaginary axis except at the origin, we see that the $z$-integral is completely regularized. It receives contributions from $z=-\eps \pm \frac{\pi i}{s+1}$ and from the points in the interior, i.e.  $z = z_* - \eps \pm \frac{\pi i}{s+1}$, where $z_*$ are zeros or poles of $G$ for which necessarily $\re z_* <0$. The $s$ integral is then convergent because of the suppression factor $e^{-s\eps}$. In the limit $\eps \to 0$ there might however arise terms $1/\eps$ if the non-exponential part of the integrand does not vanish for large $s$.

To compute such a possible anomaly in the equation of motion we set once again
\be
G(z) = \frac{1}{z^r} g(z),
\ee
assuming that $g(z)$ is holomorphic and nonvanishing at zero. For integer $r>0$ the integrand of (\ref{tF123}) has a pole of order $r$ so to compute the residue one has to perform $r-1$ derivatives. The leading term in $s$ is obtained when these act on $e^{sz}\left(z+\eps \pm\frac{\pi i}{s+1}\right)^r$. After performing the residue integral the leading behavior for large $s$ is $e^{-s\eps}\times O(1)$ giving upon $s$-integration a factor of $1/\eps$ which cancels an overall factor of $\eps$. The final answer in $\eps \to 0$ limit is a finite nonzero anomaly
\be
\aa = \frac{\pi}{4} \left( L_{r-1}^{(2)} (2\pi i) +  L_{r-1}^{(2)} (-2\pi i)\right) c_1 c_0 \ket{0} + \cdots,
\ee
where $L_n^{(\alpha)}(z)$ are generalized Laguerre polynomials. Equivalent expression which makes sense also for non-integer $r$, both positive or negative, is
\be
\aa = \frac{\pi}{8} r(r+1) \left[  {}_1 F_1(2+r,3,2\pi i)+{}_1 F_1(2+r,3,-2\pi i)\right] c_1 c_0 \ket{0} + \cdots.
\ee
This manifestly vanishes for $r=-1,0$ corresponding to the tachyon and perturbative vacuum respectively. For the double brane, i.e. $r=1$, the coefficient is equal to $\pi/2$ for all choices of $g(z)$.\footnote{For the 'ghost branes' with $r\le-2$ there is an additional $g$-dependent contribution to which the first term in $F^2= 1-G$ contributes.} Since this anomaly is independent of $g$ there is no obvious way to cancel it by a clever choice of $g$. The simplest possibility is to avoid the anomaly by taking the contour of the $z$-integration to bypass the singularity on the left. With the $\eps$-regularization this means that the contour must pass to the left of $-\eps$. As we have seen, such a prescription is very natural also when considering the kinetic term. For consistency we should therefore adopt the same prescription when we compute the coefficients. This will lead to a well-defined phantom term, which we shall study later in Sec. \ref{sec:Phantom}.  Note however, that this prescription is possible only for integer $r$, when there is no branch cut extended to the origin. The anomaly in the equations of motion for the 'fractional' D-branes is thus inevitable. Alternative strategy for canceling the anomaly would be to consider functions $G$ with an infinite number of zeros or poles along the imaginary axis. We will postpone such an attempt to the future work.

\section{Level expansion}
\label{sec:LevelExp}
\setcounter{equation}{0}

\subsection{General arguments for convergence}
\label{sec:Convergence}

As we have already explained, higher multiple D-brane solutions are associated to functions $G(K)$ that have a pole singularity at the origin, and therefore it is of utmost importance to carefully examine whether such solutions make sense. Up to date, inverse powers of $K$ have appeared in the literature in attempts to write formally the tachyon vacuum as a pure gauge configuration. It has been argued that such string fields are not to be considered as good string fields for the purpose of a gauge transformation and hence the non-triviality of the tachyon solution remains undisputed.

The reason why objects like $1/K$ are dangerous is very simple: at least formally we have
\be
\frac{1}{K} = \int_0^\infty d\alpha \, e^{-\alpha K},
\ee
and regarded as a string field this is linearly divergent since
\be
\lim_{\alpha\to\infty}  e^{-\alpha K} = \ket{\infty} = e^{ - \frac{1}{3} L_{-2} +  \frac{1}{30} L_{-4} + \cdots} \ket{0}
\ee
is the well-known sliver state \cite{RZ}. We thus have to be very careful whenever inverse powers of $K$ appear. Let us study in more detail string fields of the form
\be
c K^n B c K^m,
\ee
where $n$ and $m$ are integers, possibly negative. The coefficients of such fields in the standard Fock space basis can be easily computed by calculating the overlaps with Fock states, or more systematically using the tools developed in \cite{wedge, Schnabl} and reviewed in Appendix \ref{sec:app-level}.

The starting point is to compute the coefficients in
\be
e^{-\alpha K} c B e^{-\beta K} c e^{-\gamma K}
\ee
and obtain the coefficients of $c K^n B c K^m$ by differentiation with respect to the parameters $\beta$ and $\gamma$ and setting them to zero, or by integration, depending on the sign of $n$ and $m$.

The coefficients are given for reader's convenience in a compact form in Appendix~\ref{sec:app-level}. To write them explicitly, one has to choose a basis of states. Most convenient one for work with analytic universal solutions in $L_0$-level expansion, is the basis formed by $b$ and $c$ ghosts, and total Virasoro generators $L$ acting on the $SL(2,\rr)$ invariant vacuum $\ket{0}$.

It turns out that potentially most divergent terms are those that include only ghosts, so let us focus our discussion only on states of the form $c_{-n} \ket{0}$, ignoring all other for the moment.
Instead of listing the coefficients $t_n$ of $c_{-n} \ket{0}$ explicitly, let us write down the asymptotic behavior for large $\gamma$ with fixed $\beta$ (setting $\alpha=0$)\footnote{The expansion for large $\beta$ and constant $\gamma$ is needed for the discussion of 'ghost branes'. The expansion starts at order $\beta^{-3}$ with coefficients which are polynomial or rational functions of $\gamma$.}
\bea
t_{-1} &=&\frac{\pi}{8} - \frac{\pi^3}{96 \gamma^2} + \frac{(1  - 4  \beta^2 -4  \beta^3) \pi^3 }{48 \gamma^3} + \oo(\gamma^{-4})
\nonumber\\
t_0 &=&\frac{1}{2} - \frac{\pi^2}{12 \gamma^2} + \frac{(1 - 3 \beta^2 - 2  \beta^3) \pi^2 }{6 \gamma^3} + \oo(\gamma^{-4})
\nonumber\\
t_{1} &=&\frac{2}{\pi} - \frac{\pi}{2 \gamma^2} + \frac{(3 - 6 \beta^2 - 4  \beta^3) \pi }{3 \gamma^3} + \oo(\gamma^{-4})
\nonumber\\
t_{2} &=&\frac{8}{\pi^2} - \frac{8}{3 \gamma^2} + \frac{16(1 +2\beta - \beta^2 - 4 \beta^3 - 2  \beta^4) }{3(1+2\beta) \gamma^3} + \oo(\gamma^{-4}).
\eea
It is important to notice that all coefficients up to order $\gamma^{-2}$ are independent of $\beta$, and therefore they would vanish when a derivative with respect to $\beta$ is taken.  Starting from $t_2$ the coefficient of $\gamma^{-3}$ will develop a nontrivial denominator thanks to the presence of uncanceled tangents in the expansion of the $c$-ghost.

Inspecting closer the coefficients of $c_1 \ket{0}$, $c_0 \ket{0}$, and $c_{-1} \ket{0}$, we find that in general they are unambiguously defined in the string field $c K^n B c K^m$ whenever $m+n \ge 1$. The reason being that as long as one of the integers is positive, the respective $K$ turns effectively into a derivative and reduces thus the divergence in the integration over the wedge-width variable associated with the other power, if it is negative. Note that the criterion $m+n \ge 1$ can be nicely restated in terms of the $L^-$ weight, which is related to the $\ll_0$ weight. The nice thing about the $L^-$ weight, is that it is strictly additive in the $K, B, c$ algebra. The fields $K$ and $B$ each carry weight one, while the field $c$ has weight minus one. So we can say that the coefficients of $c_1 \ket{0}$, $c_0 \ket{0}$, and $c_{-1} \ket{0}$ in $c K^n B c K^m$ are finite as long as the total $L^-$ weight is greater or equal to zero.\footnote{There are nine exceptions to this rule, the states $cBc$, $cKBc K^{-n} \,(n=1,2)$, $cK^2 Bc K^{-2}$, $cB K^{-1} c K^n\, (n=0,1)$, and $cBK^{-2} c K^n$ with $n=0,1,2$. Although the $L^-$ weights of states are $-1$, $-2$, or $-3$, they have well defined absolutely convergent coefficients in front of $c_1 \ket{0}$, $c_0 \ket{0}$, and $c_{-1} \ket{0}$.}

On the other hand, coefficients of higher level states such as $c_{-2}\ket{0}$ or $c_{-3}\ket{0}$  might be divergent even when the condition $m+n \ge 1$ is satisfied. The reason is that for these coefficients repeated derivative with respect to $\beta$ does not improve the large $\gamma$ behavior. Therefore these coefficients are divergent for  $m \le -3$ regardless of the value of $n$. Similarly, they are divergent for $n \le -3$ regardless of the value of $m$.

The divergences can be alternatively seen by computing overlaps with the Fock states. For the potentially most divergent piece $c K^{n+1} B c K^{-n}$ in the multibrane solution we find
\bea
\aver{\phi \left|c K^{n+1} B c \frac{1}{K^n}\right.} &=& \int_0^\infty dt \frac{t^{n-1}}{(n-1)!} \Tr\left(e^{-K/2} \phi\, e^{-K/2} c K^{n+1} B c \, e^{-t K} \right) \\
&=& \int_0^\infty dt \frac{t^{-h_V}}{(n-1)!} \Tr\left(e^{-\frac{K}{2t}} c\partial c V\, e^{-\frac{K}{2t}} c K^{n+1} B c \, e^{- K} \right),
\eea
where in the second line we used as an example $\phi = c\partial c V$ with $V$ being a purely matter field of dimension $h_V \ge 0$. The $c$ ghost OPE's give us a factor of $1/t^2$ so that the integral is nicely convergent at large $t$. (There is no issue at small $t$.)
Choosing however  $\phi = :c\partial c \partial^2 c b:$ we get an integral of the form $\int_0^\infty dt t^{n-4}$. Therefore the coefficient of $c_{-2} \ket{0}$ is divergent for $n \ge 3$, in agreement with our previous analysis. Similarly we would find divergence for $n \le -4$.

\subsection{Towards the phantom}
\label{sec:Phantom}

As we have just seen, the coefficients of $c K^{n+1} B c K^{-n}$ in the Fock state expansion are finite for $-3 \le n \le 2$, and such a statement remains true even if we replace the powers of $K$ with arbitrary functions of $K$, as long as we do not change the behavior near $K=0$. So we could just compute the coefficients of $F_1(K) c F_2(K) B c F_3(K)$ by using the Laplace representation of the functions $F_i(K)$ and the knowledge of the coefficients in $e^{-\alpha K} c B e^{-\beta K} c e^{-\gamma K}$. Throughout the paper we were faced with analogous computations, for which we used the so called $s$-$z$ trick number of times. We have seen on the examples of the kinetic term energy (for more than two branes), Ellwood invariant, or the anomaly in the equation of motion, that we get the correct answer, if and only if our $z$-integration contour running along the imaginary axis bypasses the singularity at the origin from the left. On the other hand, straightforward integration over $\alpha$, $\beta$ and $\gamma$ gives the same answer, as if the $z$ contour bypassed the origin on the right! For numerical computations, presented in the next subsection, it is more convenient to find the coefficients by integrating over $\beta$ and $\gamma$ (for this class of solutions $\alpha$ is simply set to zero), than by computing the integrals over $s$ and $z$. Therefore, we will now describe how to go around the singularity on the left, in terms of the $\beta, \gamma$ integrals.

General Fock state coefficient of $c F_2(K) Bc F_3(K)$ can be computed as
\be\label{bccoeff}
\int_0^\infty\!\! d\beta \int_0^\infty\!\! d\gamma\,  f_2(\beta) f(\beta, \gamma) f_3(\gamma),
\ee
where $f_2(\beta)$ and $f_3(\gamma)$ are inverse Laplace transforms of $F_2(z)$ and $F_3(z)$. The function $f(\beta, \gamma)$ denotes the selected coefficient in the string field $c B e^{-\beta K} c e^{-\gamma K}$. This expression can be rewritten in full generality as
\be\label{szcoeff}
\int_0^\infty\!\! ds \int_{-i \infty}^{i \infty} \frac{dz}{2\pi i}\, e^{s z} \left(F_2(z) f(-\leftpartial_z, s+\leftpartial_z)\right) F_3(z).
\ee
For the coefficients of $c_1 \ket{0}$, $c_0 \ket{0}$, and $c_{-1} \ket{0}$ life is simpler once again, the function $f(-\leftpartial_z, s+\leftpartial_z)$ depends on the derivative only through a linear factor, or through a simple exponential, such as $e^{\pm \frac{2\pi i}{1 + s} \leftpartial_z} $, which acts as a translation operator. For higher level coefficients such as $c_{-2}\ket{0}$ or $c_{-3}\ket{0}$ we get infinite number of exponentials (upon expansion of tangents) and this presents extra challenges discussed later.

In the formula \eqref{szcoeff} the line integral along the imaginary axis passes through a singularity at the origin (due to the presence of either $F_2$ or $F_3$), and one has to specify a prescription. The prescription which gives the same result as \eqref{bccoeff} can be found in the following way. As we have shown in the previous subsection, the integral \eqref{bccoeff} is absolutely convergent for $-3 \le n \le 2$, and for simplicity we shall restrict our attention to this range only. We can thus multiply the integrand by $e^{-(\beta+\gamma)\eps}$. The limit $\eps \to 0$ is well defined and gives back our integral. Now rewriting this $\eps$-dependent integral using the $s$-$z$ trick, we get back \eqref{szcoeff}, but with shifted arguments of the $F$ functions, i.e. $F_i(z+\eps)$ instead of $F_i(z)$. Since the contour runs along the imaginary axis and the singularity has been shifted to $-\eps$, we see that the contour bypasses it on the right. This remains so, even in the limit of vanishing~$\eps$.

Our proposal for the phantom term is to exclude or subtract the contribution from the origin. Let us demonstrate on a simple example of a regular string field
\be
c \frac{1}{K+a} Bc \frac{1}{K+b}
\ee
how we can easily separate the contributions from the residues in $-a$ and $-b$. Let us denote $F_2(z)=(z-a)^{-1}$ and $F_3(z)=(z-b)^{-1}$, and by $f_2(\beta)$ and $f_3(\gamma)$ their inverse Laplace transforms.  Assuming for definiteness $a>b>0$, we shall use for the contribution from $-b$ the identity
\bea
\left. F_2(z) f(-\leftpartial_z, s+\leftpartial_z)\right|_{z=-b} &=& \int_0^\infty d\beta\, \left. e^{-\beta z} f_2(\beta) f(\beta,s-\beta)\right|_{z=-b}
\nonumber\\
&=& \int_0^\infty d\beta\,e^{-\beta (a-b)} f(\beta,s-\beta),
\eea
while for the contribution from $-a$ we write
\bea
\label{factF}
\left. f(s+\rightpartial_z, -\rightpartial_z) F_3(z) \right|_{z=-a} &=& - \int_0^\infty d\gamma\, e^{\gamma z} f_3(-\gamma) f(s+\gamma,-\gamma)\bigr|_{z=-a}
\nonumber\\
&=& - \int_{-\infty}^0 d\gamma\,e^{-\gamma (b-a)} f(s-\gamma,\gamma).
\eea
In the first line of (\ref{factF}) we have used the Laplace representation of the function $F_3(z)$ valid in the domain $\re z< -b$, and in the second line we just renamed the integration variable from $\gamma$ to $-\gamma$. The whole coefficient with contributions from both residues is then simply
\be
\int_0^\infty ds \, e^{-bs} \int_0^\infty d\beta\,e^{-\beta (a-b)} f(\beta,s-\beta) - \int_0^\infty ds \, e^{-a s} \int_{-\infty}^0 d\gamma\,e^{-\gamma (b-a)} f(s-\gamma,\gamma).
\ee
Upon a change of variable $\gamma=s-\beta$, this can be written as
\be
\int_0^\infty ds \, e^{-bs} \left( \int_0^\infty - \int_s^\infty \right)d\beta \,e^{-\beta (a-b)} f(\beta,s-\beta),
\ee
where the first integral in the bracket gives the contribution from $-b$ while the second from $-a$. The sum of the two is
\be
\int_0^\infty ds \, e^{-bs}  \int_0^s d\beta \,e^{-\beta (a-b)} f(\beta,s-\beta) =\int_0^\infty ds \int_0^s d\beta \,f_2(\beta) f_3(s-\beta) f(\beta,s-\beta) ,
\ee
which is nothing but \eqref{bccoeff}.

So we propose a simple prescription for the phantom. From the naive solution whose coefficients are computed using the $\beta$ and $\gamma$ integrals, one should subtract a finite phantom term which is defined as the contribution from the pole of $F_3$ in the case of the double brane.
Explicitly it takes the form
\be
\mbox{Phantom} = \int_0^\infty\!\! ds \int_0^\infty\!\! d\beta\,  f_2(\beta) f(\beta, s-\beta) f_3(s-\beta).
\ee
Although, for simplicity, we have derived it here assuming that $F_2$ and $F_3$ have at most a single pole, the final answer we obtained correctly reproduces the residuum even in the case when we want to isolate the contribution from higher order poles of $F_3$ at the origin.
To conclude, a general coefficient for the double brane solution is given by
\be\label{gc-with-Phantom}
\int_0^\infty\!\! d\beta \int_0^\infty\!\! d\gamma\,  f_2(\beta) f(\beta, \gamma) f_3(\gamma) - \int_0^\infty\!\! ds \int_0^\infty\!\! d\beta\,  f_2(\beta) f(\beta, s-\beta) f_3(s-\beta).
\ee
One of the potential subtleties of this phantom, is that for certain higher level coefficients that contain higher powers of
\be
\tan\left(\frac{\pi}{2}\frac{\beta-\gamma}{1+\beta+\gamma}\right) = \tan\left(\frac{\pi}{2}\frac{2\beta-s}{1+s}\right),
\ee
the integrand becomes singular for an infinite number of values of $\beta = s+\half +(k+1)s$. Note that none of these values lies in the interval $(0,s)$, which is the range of the naive term \eqref{bccoeff} and which is therefore finite and unambiguous. When we encounter such singularities, we adopt principal value prescription. Numerically we implement it by rotating the contour of the $\beta$ integral from the positive real axis into the complex $\beta$ plane, slightly up or down, and taking the arithmetic mean of these two results. This also cancels the unwanted imaginary parts.

\subsection{Numerical results}

So far, most of our discussion of multiple D-brane solutions in previous sections was rather general, in terms of a function $G(K)$, which led us to impose some important conditions. Let us now look in more detail at a concrete example of the solution
\be\label{Nsol}
\Psi = c \frac{K^{n+1}}{(1+K)^n} B c \left(1- \frac{(1+K)^n}{K^n}\right)
\ee
for which
\be
G(z) = \left(\frac{z+1}{z}\right)^n .
\ee
This family of solutions has two well known members: for $n=0$ this is the perturbative vacuum $\Psi=0$, while for $n=-1$ it is the simple tachyon vacuum solution of \cite{ErS} .
If the tachyon vacuum were a large gauge transformation of the perturbative vacuum (which it is not in any good sense), this whole family could be viewed as given by the powers of the tachyon gauge transformation.

It is quite straightforward to compute any coefficient in the expansion of $\Psi$. For example for $n=1$, which is a candidate for the double brane, we find the tachyon coefficient of $c_1 \ket{0}$ to be
\bdm
t=0.372994 - 0.588638 = -0.215644.
\edm
The first term is the naive contribution, the second term subtracts the residue at zero.

In the following tables we list coefficients from $c_1 \ket{0}$ to  $c_{-5} \ket{0}$ for values of $n$ between $n=2$ (triple brane) to $n=-3$ (ghost double brane). The first table includes only the naive or regular term.
\begin{center}
\begin{tabular}{|r|r|r|r|r|r|r|r|}
\hline
$n$  & $c_1 \quad$      & $c_0\quad$ & $c_{-1}\quad$ & $c_{-2}\quad$ & $c_{-3}\quad$ & $c_{-4}\quad$ & $c_{-5}\quad$ \\
\hline
$2$  & $0.53730$   & $0.77577$  & $0.48429$  & $0.03048$  & $-0.70900$ & $-1.73790$  & $-3.20121$  \\
\hline
$1$  & $0.37299$  & $0.44803$  & $0.45303$  & $0.47698$  & $0.49104$  & $0.51026$  & $0.51867$   \\
\hline
$-1$ & $0.28439$  & $0.24903$  & $0.24452$  & $0.25257$  & $0.26894$  & $0.29239$  & $0.32283$   \\
\hline
$-2$ & $0.63671$  & $0.23069$  & $0.16044$  & $-0.03167$ & $-0.13167$ & $-0.33494$ & $-0.50203$  \\
\hline
$-3$ & $1.37982$  & $0.09646$  & $0.41052$  & $-0.31435$ & $-0.13341$ & $-0.75982$ & $-0.68690$  \\
\hline
\end{tabular}
\end{center}
The second gives the final coefficient with the phantom subtracted (i.e. included).
\begin{center}
\begin{tabular}{|r|r|r|r|r|r|r|r|}
\hline
$n$  & $c_1 \quad$      & $c_0\quad$ & $c_{-1}\quad$ & $c_{-2}\quad$ & $c_{-3}\quad$ & $c_{-4}\quad$ & $c_{-5}\quad$ \\
\hline
$2$  & $-0.43481$ & $-0.18127$ & $0.02353$  & $-0.32736$ & $-1.20906$ & $1.45109$  & $4.13767$  \\
\hline
$1$  & $-0.21564$ & $-0.06671$ & $0.00911$  & $-0.08671$ & $-0.49105$ & $0.25643$  & $1.88705$  \\
\hline
$-1$ & $0.28439$  & $0.24903$  & $0.24452$  & $0.25257$  & $0.26894$  & $0.29239$  & $0.32283$  \\
\hline
$-2$ & $0.64199$  & $0.48567$  & $0.40056$  & $0.35881$  & $0.70399$  & $0.93730$  & $-0.91004$ \\
\hline
$-3$ & $0.97760$  & $0.64625$  & $0.48330$  & $0.30180$  & $0.92385$  & $2.27210$  & $-1.31707$  \\
\hline
\end{tabular}
\end{center}
The results for the ghost branes with $n<-1$ were obtained by a variant of the formula (\ref{gc-with-Phantom}), where the integration variable in the second term is changed from $\beta$ to $\gamma$, and correspondingly in the integrand $\beta$ is replaced by $s-\gamma$.

How can we tell whether the results are meaningful? First thing which may worry the reader is that the coefficients do not decay rapidly, in fact, they do not decay at all at higher levels. But this is a well-known feature of the non-twist invariant solutions, such as the ES solution \cite{ErS} given on the third line. Our solution \eqref{Nsol} contains $c$ on the left, and therefore it must be annihilated by the operator $c(1)$.

Another very positive aspect we can read from the table relates to the Ellwood invariant.
The Ellwood invariant measures the change in the closed string one point function and it can be easily computed using the conservation laws found in \cite{KKT, Kishimoto}. The laws allow one to reduce the computation of $\aver{I|c\bar c V(i)|\phi}$ for arbitrary $\phi$ in the universal sector to the case of $\ket{\phi}= c_1 \ket{0}$. For a string field of the form
\bea
&& t_1 c_1 \ket{0}  + t_2 b_{-2} c_0 c_1  \ket{0} + t_3 c_{-1} \ket{0} + t_4 L_{-2} c_1 \ket{0} + t_5 b_{-2} c_{-1} c_0 \ket{0} + t_6 b_{-2} c_{-2} c_1 \ket{0}+ t_7 b_{-4} c_{0} c_1 \ket{0} \nonumber\\
&&  \quad + t_8 c_{-3} \ket{0} + t_{9} L_{-2} b_{-2} c_0  c_1 \ket{0}+ t_{10} L_{-2} c_{-1} \ket{0} + t_{11} L_{-4} c_{1} \ket{0} + t_{12} L_{-2}L_{-2} c_{1} \ket{0}  \cdots
\eea
the Ellwood invariant, normalized so that for the tachyon vacuum it gives $+1$, is given by
\be
\frac{\pi}{2} \left( t_1 + t_3 + t_6 + 4 t_9 -3 t_{10} -4 t_{12}  + \cdots \right).
\ee
The odd level fields (with even eigenvalue of $L_0$) do not contribute, but many other terms do not contribute as well. This formula explains why for the (especially twist invariant) tachyon vacuum solutions, known to be behaving well in level truncation, the tachyon coefficients are close to $2/\pi$. Incidentally, in the $\ll_0$-level truncation, the tachyon coefficient of the $\bb_0$-gauge solution \cite{Schnabl} has exactly this value.

Looking at the second table, we see that the $c_1$ coefficient depends on $n$ roughly in a linear manner, and that for each conjectured $n+1$ brane solution, the contribution from this coefficient gives between 34\% to 51\% of the expected Ellwood invariant. This is quite encouraging, especially in comparison to the tachyon vacuum solution, for which the $c_1$ coefficient accounts for about 45\% of the expected value, and we know from other studies that it behaves well in many respects.

For more detailed analysis we have selected the double brane solution with the phantom. From the Fock state coefficients up to level 16 we have found the kinetic term and the Ellwood invariant. We give our results in the form of a polynomial in $z$, in which the variable $z$ acts as a level counting variable. For the kinetic term, normalized so that the expected value would be $+1$ at $z=1$ we found
\bea
E &=& -0.152973 - 0.0292659 z + 0.260087 z^2 + 0.0787223 z^3 -
 1.84353 z^4 + 0.264623 z^5
 \nonumber\\
 && - 1.2558 z^6 + 2.46205 z^7 -
 26.0792 z^8 + 14.8247 z^9 - 83.7969 z^{10} + 81.068 z^{11}
 \nonumber\\
 &&  -
 450.531 z^{12} + 387.393 z^{13} - 1731.8 z^{14} + 1673.6 z^{15} -
 6957.27 z^{16} + \cdots \,.
\eea
Following \cite{Schnabl, ErS} one may attempt to resum this apparently divergent series using Pad\'{e} approximants. We will not present the results in detail here, since we did not find them illuminating enough. We think there is more to understand, than what we have been able so far.
Anyway, our observations from the Pad\'{e} analysis are that things look promising at lower levels. The energy has to start negative (i.e. with the wrong sign), on general grounds, so is reassuring to see that the value $-0.15$ is not too far from zero. Also the second correction which has to be negative is rather small. Then come finally two values which give positive contribution. Up to this level, everything looks quite nice. But then the positive contributions which come mostly at even levels (odd powers of $z$) seem to be smaller than the negative contributions from both neighboring levels. This seems to imply the wrong sign of the resummed answer, but the change in the expected value from level to level is too high to draw any definite conclusion.

Similarly, for the Ellwood invariant, normalized so that the expected value for the double brane is $-2/\pi$ we have obtained
\bea
\widetilde E &=&-0.215634 + 0.0091296 z^2 + 0.184435 z^4 + 0.205632 z^6 -
 1.23245 z^8 + 1.38278 z^{10}
 \nonumber\\
 &&
 - 1.7683 z^{12} + 5.19292 z^{14} - 11.6944 z^{16} + \cdots \, .
\eea
Again, the Pad\'{e} analysis does not provide much help, so we do not present the details here. Optically it seems to provide about one third of the expected answer (with the correct sign), but the variability is again very high.
For comparison, let us show the analogous polynomial computed to the same level, in the same normalization, for the asymmetric simple solution \cite{ErS} for the tachyon vacuum
\bea
\widetilde E_{TV} &=& 0.284394 + 0.244516 z^2 - 0.0375272 z^4 + 0.0504927 z^6 +
 0.0925147 z^8 - 0.0026853 z^{10}
 \nonumber\\
 &&
  - 0.105459 z^{12} + 0.00828519 z^{14} +
 0.148443 z^{16}\cdots \, .
\eea
In this case the convergence of the Pad\'{e} approximants at $z=1$ to the expected value is quite apparent, and up to this level one finds results within about 7\% of the expected answer.

What should we conclude from these numerical results? One possibility is that we have not gone to high enough levels to see the convergence, or that we have not computed our numerical coefficients accurately enough. To this level we had to compute about 1316 coefficients, each given by quite a slowly convergent double integral. Another possibility is that our prescription for computing certain coefficients, such as the one of $c_{-3} \ket{0}$ for which the phantom naively diverges, is incorrect. Third possibility is that because of the specific analytic structure of the energy, as a function of the level counting parameter $z$, the series is simply not Pad\'{e} summable, and that in the tachyon vacuum case where it did work, we were just lucky. More serious possibility, of course, is that we have identified the phantom incorrectly.

\section{Discussion}
\label{Discussion}
\setcounter{equation}{0}

In this work we have studied Okawa's family of string field theory solutions depending on a single analytic function. We have shown how to compute their energy and the closed string one-point functions, and that these two computations agree. Our results immediately suggest that functions with a pole at the origin should be interpreted as describing multiple D-branes.

One question one could ask is whether our solutions favor multiple D-branes over unphysical
configurations with negative number of D-branes. They do, but only a little bit. For negative $n$,
the solution \eqref{Nsol} contains a piece with negative weight. Such states are generally
problematic in correlators and in the overlaps with Fock states. But we have not seen anything
wrong with the one and two ghost brane solutions. We expect that a more detailed analysis would
rule out such solutions. Another question is related to the existence of fractional number of
D-branes. We certainly do not expect their existence in the universal sector of bosonic string
field theory. Such solutions are plagued by irreparable anomalies as we have shown in
Sec.~\ref{sec:Anomalies}.

We have seen that our proposed multibrane solutions have to come with a special prescription, of
bypassing a singularity at the origin from the left. We have interpreted such a prescription as a
sort of finite phantom term, but this is not a unique possibility. In Sec.~\ref{sec:Anomalies} we
have seen another type of phantom which arises when we systematically replace all $K$'s with
$K+\eps$. The latter phantom vanishes in level truncation, but at the same time it provides the
sole contribution to the Ellwood invariant. We have tested the proposed double brane solution
numerically, but the results we got are inconclusive. Settling this issue is postponed for a future
work. It would be very nice if more regular solutions, which would behave well in level truncation,
were found.

\section*{Note added in proof}

The referee of this paper wishes to comment that the term ${\cal A}_\eps$, introduced in
section~\ref{s_anom_eom}, in the limit $\eps \to 0$ is a distribution-like object, in that it has
support only on the zero mode of $K_1^L$. As a consequence a reliable   mathematical treatment
cannot be based on simple, however scrupulous, algebraic manipulations, but it must be based on
identifying the space of dual test states and evaluating ${\cal A}_\eps$ against them.

\medskip
\noindent Following upon the referee's remark, the authors observed that $\lim_{\eps\to 0} ({\cal
A}_\varepsilon K^n)$ indeed vanishes in the Fock space for the multiple D-brane solution, for $n$
greater or equal to the degree of the pole of $G$ at the origin, i.e. the number of D-branes minus
one. This is analogous to the statement that $\delta^{(m)}(x)\, x^n =0$ for $n \ge m+1$. Developing
rigorous distribution-theory-like framework for string field theory, however, presently seems to be
a rather challenging task.

\section*{Acknowledgments}

\noindent
 We would like to thank Ian Ellwood, Ted Erler, Koji Hashimoto, Hiroyuki Hata, Toshiko Kojita, Carlo Maccaferri, Toru Masuda, Yuji Okawa, Ashoke Sen and Daisuke Takahashi for useful discussions.
The work of M.M. (No.~21-173) was supported by Grants-in-Aid for Japan
Society for the Promotion of Science (JSPS) Fellows.
M.M was also supported by the Grant-in-Aid for the Global COE Program
"The Next Generation of Physics, Spun from Universality and Emergence"
from the Ministry of Education, Culture, Sports, Science and
Technology (MEXT) of Japan. M.S. would like to thank Aspen Center for Physics with their NSF grant No.~1066293 and to Centro de Ciencias de Benasque Pedro Pascual for providing a stimulating environment during various stages of this project. We gratefully acknowledge a travel support from a joint JSPS - M\v{S}MT grant LH11106.
The research of M.S. was supported by the EURYI grant GACR EYI/07/E010 from EUROHORC and ESF.

\appendix
\section{Coefficients in level expansion}
\label{sec:app-level}
\setcounter{equation}{0}

In this short appendix we would like to remind the reader some results from \cite{wedge,Schnabl} and show how to efficiently compute coefficients in the level expansion of
\be\label{-cB-c-}
e^{-\alpha K} c B e^{-\beta K} c e^{-\gamma K}.
\ee
This can be written as a state in the Hilbert space
\be
\mscriptsize{\frac{1}{\pi}} \widehat U_{\alpha+\beta+\gamma+1} \left[ \tilde c \mscriptsize{\left(\frac{\pi}{2} \frac{-\alpha+\beta+\gamma}{\alpha+\beta+\gamma+1}\right)} + \tilde c \mscriptsize{\left(\frac{\pi}{2} \frac{-\alpha-\beta+\gamma}{\alpha+\beta+\gamma+1}\right)}
-\mscriptsize{\frac{2}{\pi}}  \widehat\bb \, \tilde c \mscriptsize{\left(\frac{\pi}{2} \frac{-\alpha+\beta+\gamma}{\alpha+\beta+\gamma+1}\right)} \tilde c \mscriptsize{\left(\frac{\pi}{2} \frac{-\alpha-\beta+\gamma}{\alpha+\beta+\gamma+1}\right)} \right] \ket{0},
\ee
where $\widehat U_r \equiv U_r^\star U_r$, $\widehat\bb = \bb_0 + \bb_0^\star$, the star denotes BPZ conjugation, and the remaining symbols follow the notation of \cite{Schnabl}. In particular
\bea
\tilde c(x) &=& \cos(x)^2 c(\tan x), \\
U_r &=& \left(\frac{2}{r}\right)^{\ll_0} =  \left(\frac{2}{r}\right)^{L_0} e^{u_2 L_2} e^{u_4 L_4} \ldots,
\eea
where $u_n$ are constants given in \cite{Schnabl}.
 More convenient form of the string field \eqref{-cB-c-} for the purposes of level expansion is a 'normal ordered' form
\bea
&&
\msmall{\frac{1}{\pi}} U_{\alpha+\beta+\gamma+1}^\star \left[ (\msmall{\gamma+\half})\tilde c \msmall{\left(\frac{\pi}{2} \frac{-\alpha+\beta+\gamma}{\alpha+\beta+\gamma+1}\right)} + (\msmall{\alpha+\half}) \tilde c \msmall{\left(\frac{\pi}{2} \frac{-\alpha-\beta+\gamma}{\alpha+\beta+\gamma+1}\right)} \right] \ket{0}
\nonumber\\
&&
-\msmall{\frac{\alpha+\beta+\gamma+1}{\pi^2}} U_{\alpha+\beta+\gamma+1}^\star  \bb_0^\star  \, \tilde c \msmall{\left(\frac{\pi}{2} \frac{-\alpha+\beta+\gamma}{\alpha+\beta+\gamma+1}\right)} \tilde c \msmall{\left(\frac{\pi}{2} \frac{-\alpha-\beta+\gamma}{\alpha+\beta+\gamma+1}\right)} \ket{0}.
\eea
For example the coefficient of $c_1 \ket{0}$ can be easily read off
\bea
&&
 \mscriptsize{\frac{\alpha+\beta+\gamma+1}{2\pi}} \left[ (\mscriptsize{\gamma+\half})\cos^2 \mscriptsize{\left(\frac{\pi}{2} \frac{-\alpha+\beta+\gamma}{\alpha+\beta+\gamma+1}\right)} + (\mscriptsize{\alpha+\half}) \cos^2 \mscriptsize{\left(\frac{\pi}{2} \frac{-\alpha-\beta+\gamma}{\alpha+\beta+\gamma+1}\right)} \right]
\\\nonumber
&&
-\mscriptsize{\frac{(\alpha+\beta+\gamma+1)^2}{2\pi^2}}  \left(\tan\mscriptsize{\left(\frac{\pi}{2} \frac{-\alpha+\beta+\gamma}{\alpha+\beta+\gamma+1}\right)} -\tan\mscriptsize{\left(\frac{\pi}{2} \frac{-\alpha-\beta+\gamma}{\alpha+\beta+\gamma+1}\right)} \right) \cos^2\mscriptsize{\left(\frac{\pi}{2} \frac{-\alpha+\beta+\gamma}{\alpha+\beta+\gamma+1}\right)} \cos^2\mscriptsize{\left(\frac{\pi}{2} \frac{-\alpha-\beta+\gamma}{\alpha+\beta+\gamma+1}\right)}.
\eea

\section{What is behind the $s$-$z$ trick?}
\label{sec:app-sz}

Undoubtedly much of our discussion in this paper relied on the not so transparent $s$-$z$ trick. In this appendix we will try to clarify it a bit in a simpler setting.
Instead of looking at complicated correlators of strings fields with ghost insertions, let us apply the $s$-$z$ trick to a simple product of functions. Following exactly the same steps as before we get
\be
\prod_{i=1}^n F_i(z) = \int_0^\infty ds \int_{-i\infty}^{i\infty} \frac{dw}{2\pi i}\, e^{s w} \prod_{i=1}^n F_i(z+w).
\ee
In particular, for a single function $F(z)$ we find
\be
F(z) = \int_0^\infty ds \int_{-i\infty}^{i\infty} \frac{dw}{2\pi i}\, e^{s w} F(z+w).
\ee
What are the conditions of validity of such an expression? Let us look at an instructive example of
\be
F(z)=\frac{1}{1+z}.
\ee
The integration contour along the imaginary axis can be closed by adding a non-contributing arch at infinity in the $\re w <0$ half-plane. This contour integral can then be nonzero only if it encircles the pole at $w=-(z+1)$.
Therefore the right hand side equals to
\be
F_{\mbox{\scriptsize $s$-$z$}}(z) = \frac{1}{1+z}\,  \theta\left(\re(z+1)\right),
\ee
where $\theta$ is the usual Heaviside step function. This agrees with the original function $F(z)$ clearly only in the half plane $\re z>-1$. For more general rational functions $F(z)$ the corresponding $F_{\mbox{\scriptsize $s$-$z$}}(z)$ can be defined analogously using the partial fraction decomposition.

The function $F_{\mbox{\scriptsize $s$-$z$}}(z)$ is clearly not holomorphic (although it can be analytically continued), but this appears to be more of a virtue in cases where the argument is $K$ and we have to compute correlators.


\end{document}